\documentclass[twocolumn,aps,pra,amsmath,superscriptaddress,notitlepage,longbibliography]{revtex4-1}
\usepackage{amsmath,amssymb,amsfonts,bm}
\usepackage{graphicx}
\usepackage{dcolumn}
\usepackage{mathrsfs}
\usepackage{bbold}
\usepackage{dsfont}
\usepackage{dcolumn}
\usepackage[colorlinks=true,linkcolor=blue,citecolor=blue, urlcolor=blue,bookmarks=false]{hyperref}
\usepackage{changes}

\begin{document}


\title{Optical conductivity signature of Van Hove singularity in altermagnetic topological systems}

\author{Fang Qin}
\email{qinfang@just.edu.cn}
\affiliation{School of Science, Jiangsu University of Science and Technology, Zhenjiang, Jiangsu 212100, China}

\author{Rui Chen}
\affiliation{Department of Physics, Hubei University, Wuhan, Hubei 430062, China}

\author{Xiao-Bin Qiang}
\email{12331032@mail.sustech.edu.cn}
\affiliation{State Key Laboratory of Quantum Functional Materials, Department of Physics, and Guangdong Basic Research Center of Excellence for Quantum Science, Southern University of Science and Technology (SUSTech), Shenzhen 518055, China}
\affiliation{Division of Physics and Applied Physics, School of Physical and Mathematical Sciences, Nanyang Technological University, 21 Nanyang Link, 637371, Singapore}

\begin{abstract}
We investigate the topological phases, joint density of states (JDOS), optical conductivities, and magneto-optical responses of a two-dimensional $d$-wave altermagnet with spin-orbit coupling and Zeeman splitting. The system hosts gapped Dirac points at the high-symmetry points $\Gamma$, $\textrm{M}$, $\textrm{X}$, and $\textrm{Y}$. We show that the JDOS exhibits kinks at the corresponding Dirac gap frequencies and pronounced peaks at Van Hove singularities, whose positions can be tuned by the altermagnetic order. These features are reflected in the optical conductivities, with complementary signatures in their real and imaginary parts. In particular, the Van Hove signatures in the transverse optical conductivity disappear in the absence of $d$-wave altermagnetism, revealing an altermagnet-induced optical signature of the Van Hove singularity. Finally, the Faraday and Kerr rotations exhibit characteristic features inherited from the optical conductivity. Our results establish optical and magneto-optical spectroscopy as sensitive probes of Dirac gaps and Van Hove singularities in altermagnetic topological systems.
\end{abstract}

\maketitle

\section{Introduction}

Recent advances have revealed a new class of magnetic materials known as altermagnets, that host nonrelativistic momentum-space spin splitting despite their zero net magnetization~\cite{hayami2019momentum,ma2021multifunctional,smejkal2022beyond,smejkal2022emerging,mazin2022editorial,bai2024altermagnetism}. The emergence of altermagnetism is rooted in a generalized symmetry description, where symmetry operations act differently on spin and lattice, distinguishing altermagnets from conventional ferro- and antiferromagnets~\cite{smejkal2022beyond,smejkal2022emerging,mazin2022editorial,bai2024altermagnetism}. By combining the advantages of both magnetic orders, altermagnets exhibit strong spin-polarized electronic bands while avoiding stray fields~\cite{song2025altermagnets,xu2026chemical,hu2025catalog}. Experimental investigations have established $d$-wave altermagnetism in KV$_2$Se$_2$O~\cite{jiang2025metallic,wang2025atomic,yang2026visualizing} and $g$-wave altermagnetism in MnTe~\cite{krempasky2024altermagnetic,lee2024broken,osumi2024observation,orlova2025magnetocaloric} and CrSb~\cite{reimers2024direct,ding2024large,zhou2025manipulation}. Meanwhile, extensive first-principles studies have predicted a broad range of candidate altermagnets, including MnF$_2$~\cite{bhowal2024ferroically}, Mn$_5$Si$_3$~\cite{reichlova2024observation}, FeSb$_2$~\cite{mazin2021prediction,phillips2025electronic}, RbV$_2$Te$_2$O~\cite{zhang2025crystal,hu2026observation}, CsV$_2$Te$_2$O~\cite{liu2025physical}, V$_2$SeTeO~\cite{zhu2023multipiezo,marfoua2025strain}, and BiFeO$_3$~\cite{urru2025g,george2026topological,sajid2026signatures,gui2026electric}.

The momentum-space anisotropy of spin splitting and the resulting spin-polarized Fermi surfaces endow altermagnets with rich functionalities beyond conventional magnetic orders. A wide range of transport phenomena have been predicted and observed, including anomalous Hall~\cite{feng2022anomalous,tschirner2023saturation,chen2025probing}, nonlinear Hall~\cite{ezawa2024intrinsic,fang2024quantum,liu2025enhancement}, layer Hall~\cite{qin2026layer}, Josephson~\cite{ouassou2023dc,zhang2024finite,cheng2024orientation,beenakker2023phase,lu2024varphi,sun2025tunable,fukaya2025josephson,pal2025josephson}, and thermoelectric and thermal Hall effects~\cite{qin2026anomalous}. Furthermore, the interplay between altermagnetic symmetry and band topology opens new opportunities for realizing diverse topological phases, including Chern insulators~\cite{ezawa2024detecting,ma2024altermagnetic,fernandes2024topological,li2025floating,antonenko2025mirror,qu2025altermagnetic,parshukov2025topological,gonzalez2025spin}, higher-order topological phases~\cite{li2024creation,huo2026altermagnetism}, and Floquet topological phases~\cite{zhu2025floquet_Chen,liu2026linearly,ganguli2026tunable,ghorashi2025dynamical,cheraghchi2026floquet1,yarmohammadi2026floquet}.

Optical conductivity provides a versatile spectroscopic approach for revealing resonance gaps, band inversions, and singular structures in quantum materials, and has been extensively investigated both theoretically and experimentally~\cite{tse2010giant,tse2011magneto,lei2023kerr,kargarian2015theory,barati2017optical,singh2018nonlinear,hu2022signature,steiner2017anomalous,sonowal2019giant,mojarro2021optical,wang2021anomalous,carbotte2019signatures,xiong2023optical}. A pioneering work by Tse and MacDonald showed that finite-frequency optical conductivity in thin-film topological insulators can be directly detected through magneto-optical measurements, i.e., Faraday and Kerr rotations~\cite{tse2010giant,tse2011magneto,lei2023kerr}. This approach has since been extended to explore Weyl semimetals~\cite{kargarian2015theory}, nodal-line semimetals~\cite{barati2017optical}, Dirac systems~\cite{singh2018nonlinear,hu2022signature}, type-I and type-II Weyl semimetals~\cite{steiner2017anomalous,sonowal2019giant}, tilted Dirac systems~\cite{mojarro2021optical}, tilted nodal-line semimetals~\cite{wang2021anomalous}, type-I semi-Dirac materials~\cite{carbotte2019signatures}, and type-II semi-Dirac materials~\cite{xiong2023optical}, where optical responses provide valuable insights into their unconventional electronic structures.

Recently, optical conductivity has attracted considerable attention in altermagnets~\cite{chen2025magneto,Iguchi2025magneto,li2025unconventional,luo2026symmetry,rao2024tunable,zhang2025modulation,hodt2026phonon}. Previous studies have explored several aspects, including magneto-optical conductivity in the presence of discrete Landau levels~\cite{chen2025magneto,Iguchi2025magneto,li2025unconventional,luo2026symmetry}, tunable optical conductivity in altermagnets with substrate-induced spin-orbit coupling under external magnetic fields~\cite{rao2024tunable,zhang2025modulation}, and the effects of phonon scattering on the optical response of altermagnets~\cite{hodt2026phonon}. However, the optical-conductivity signatures associated with Van Hove singularities in altermagnets remain largely unexplored. In particular, how the interplay among altermagnetic order, spin-orbit coupling, and Zeeman splitting influences the Van Hove singularity features in the optical conductivity of altermagnetic topological systems has not yet been clarified.

In this work, we investigate the electronic, optical, and magneto-optical properties of a two-dimensional $d$-wave altermagnet with spin-orbit coupling and Zeeman splitting. The system hosts gapped Dirac points at the high-symmetry points $\Gamma$, $\textrm{M}$, $\textrm{X}$, and $\textrm{Y}$. The joint density of states (JDOS) exhibits kinks at the corresponding Dirac gap energies and pronounced peaks at Van Hove singularities, whose positions can be tuned by the altermagnetic order. These features are inherited by the optical conductivities, with complementary signatures in their real and imaginary parts. In particular, the Van Hove signatures in the transverse optical conductivity disappear when the $d$-wave altermagnetic term is absent, revealing an altermagnet-induced optical fingerprint of the Van Hove singularity. The corresponding Faraday and Kerr rotations further reflect the characteristic features of the optical conductivity. Our results establish optical and magneto-optical spectroscopy as sensitive probes of Dirac gaps, Van Hove singularities, and altermagnetic order in topological systems.

\section{Model}\label{2}

The system is described by the following tight-binding Hamiltonian
\begin{eqnarray}
\hat{\cal H}({\bf k}) \!=\! \hat{\cal H}_{\rm soc}({\bf k}) \!+\! \hat{\cal H}_{J}^{}({\bf k}) \!+\! \hat{\cal H}_{z},
\label{eq:Hk}
\end{eqnarray}
where ${\bf k}\!=\!(k_x,k_y)$ and the three contributions are given by
\begin{equation}
\left\{
\begin{aligned}
&\hat{\cal H}_{\rm soc}({\bf k}) \!=\! t\sin(k_{y}a)\sigma_{x} \!-\! t\sin(k_{x}a)\sigma_{y}, \\
&\hat{\cal H}_{J}^{}({\bf k}) \!=\! 2t_{J}^{}\left[\cos(k_{x}a) \!-\! \cos(k_{y}a)\right]\sigma_{z}, \\
&\hat{\cal H}_{z} \!=\! \Delta_{z}\sigma_{z}. 
\end{aligned}
\right.
\end{equation}
Here, $(\sigma_{x},\sigma_{y},\sigma_{z})$ are Pauli matrices acting on the spin degree of freedom, and $a$ denotes the lattice constant.
The first term, $\hat{\cal H}_{\rm soc}({\bf k})$, describes the Rashba spin-orbit coupling, with the parameter $t$ characterizing its strength.
The second term, $\hat{\cal H}_{J}^{}({\bf k})$, represents the $d$-wave altermagnetic order~\cite{smejkal2022beyond,smejkal2022emerging}. This term gives rise to a momentum-dependent spin splitting with vanishing net magnetization. It breaks time-reversal symmetry while preserving the combined $\hat{\cal C}_{4z}\hat{\cal T}$ symmetry~\cite{zhu2025floquet_Chen}. The parameter $t_{J}^{}$ characterizes the strength of the altermagnetic order.
The third term, $\hat{\cal H}_{z}$, describes a Zeeman-type spin splitting arising from the exchange interaction between itinerant electrons and magnetic dopants~\cite{liu2010model,zyuzin2020plane}. Physically, this term can be induced by magnetic moments aligned by the applied out-of-plane or perpendicular magnetic field~\cite{chen2025probing,liu2013in,fu2009hexagonal}.
The detailed derivation of the tight-binding Hamiltonian from the continuous Hamiltonian is provided in Sections SI and SII of the Supplemental Material~\cite{SuppMat}.

\begin{figure}[htpb]
\centering
\includegraphics[width=1.0\columnwidth]{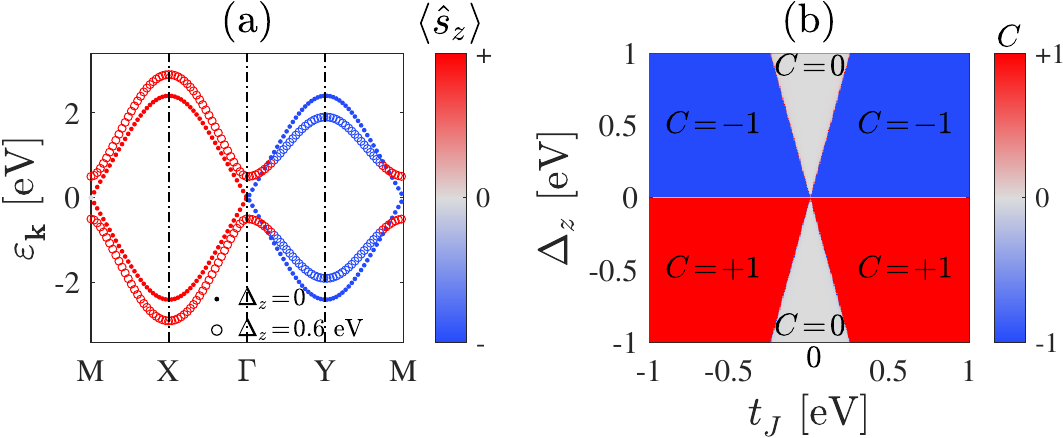}
\caption{(a) Band structure [Eq.~\eqref{eq:Ek}] for $\Delta_{z}\!=\!0$ (thin solid-circle line) and $\Delta_{z}\!=\!-0.5$ eV (thick open-circle line) at $t_{J}^{}\!=\!0.6$ eV. The color indicates the normalized expectation value of the spin operator, $\langle\hat{s}_{z}\rangle$. (b) Topological phase diagram. The Chern number $C$ [Eq.~\eqref{eq:C}] of the valence band is calculated in the gapped regime at $E_{F}\!=\!0$ as a function of the altermagnetic coupling $t_{J}^{}$ and Zeeman splitting $\Delta_z$. The red, blue, and gray regions correspond to $C\!=\!+1$, $C\!=\!-1$, and $C\!=\!0$, respectively. Here, $E_{F}$ is the Fermi energy. The other parameters are $t\!=\!1$ eV and $a\!=\!1$ nm.} \label{fig:phase}
\end{figure}

The eigenvalues for the Hamiltonian~\eqref{eq:Hk} are given by
\begin{eqnarray}
\varepsilon_{\pm\bf k}^{}\!=\! \pm\! \sqrt{\left[t\sin(k_{x}a)\right]^{2} \!+\! \left[t\sin(k_{y}a)\right]^{2} \!+\! J_{z}^{2}}, \label{eq:Ek}
\end{eqnarray} where $J_{z}^{}\!=\!2t_{J}^{}\left[\cos(k_{x}a) \!-\! \cos(k_{y}a)\right] \!+\! \Delta_{z}$, the symbol ``$\pm$'' corresponds to the conduction and valence bands, respectively. The corresponding energy band structure is shown in Fig.~\ref{fig:phase}(a) for finite altermagnetic order and spin-orbit coupling. For vanishing $\Delta_{z}$, the system remains gapless, with the band gap closing at the $\Gamma$ and $\mathrm{M}$ points, as indicated by the thin solid-circle line. In contrast, for finite $\Delta_{z}$, the system develops four gapped Dirac points centered at the four high-symmetry points, $\Gamma$, $\mathrm{M}$, $\mathrm{X}$, and $\mathrm{Y}$, as indicated by the thick open-circle line. The color scale represents the normalized expectation value of the spin operator, $\langle\hat{s}_{z}\rangle$. The anisotropic distribution of the normalized spin expectation values reflects the momentum-dependent spin splitting induced by the altermagnetic coupling.

\section{Topological phases}\label{3}

In this section, we investigate the topological phases of the model Hamiltonian in Eq.~\eqref{eq:Hk}.

\subsection{Analytical analysis}\label{3.1}

To analytically determine the topological properties of the Hamiltonian in Eq.~\eqref{eq:Hk}, we first examine the conditions under which the energy gap closes. The energy gap between the two bands is given by 
$\Delta_{g}\!=\! 2\sqrt{d_{x}^{2}\!+\!d_{y}^{2}\!+\!d_{z}^{2}}$, where $d_{x} \!=\! t\sin(k_{y}a)$, $d_{y} \!=\! -t\sin(k_{x}a)$, and $d_{z} \!=\! 2t_{J}^{}\left[\cos(k_{x}a) \!-\! \cos(k_{y}a)\right] \!+\! \Delta_{z}$. 
At the high-symmetry points $\Gamma(0,0)$, $\textrm{M}(\pi/a,\pi/a)$, $\textrm{X}(\pi/a,0)$, and $\textrm{Y}(0,\pi/a)$, the Rashba spin-orbit coupling term $\hat{\cal H}_{\rm soc}$ vanishes, such that $d_{x}\!=\!d_{y}\!=\!0$. The gap $\Delta_{g}$ therefore closes when the remaining component $d_{z}$ vanishes at any of these points. Specifically, the corresponding values of $d_z$ are $d_z(\Gamma)\!=\!d_z(\textrm{M})\!=\!\Delta_z$, $d_z(\textrm{X})\!=\!\Delta_z\!+\!4t_J$, and $d_z(\textrm{Y})\!=\!\Delta_z\!-\!4t_J$. Thus, the gap-closing points are determined by $\Delta_z\!=\!0$ and $\Delta_z\!=\!\pm4t_J$.

The Chern number of the lower (occupied) band can be obtained by summing the half-integer contributions from the four gapped Dirac points~\cite{cheraghchi2026floquet1}
\begin{eqnarray}
C\!=\!\frac{1}{2}\!\left[ {\rm sign}(\Delta_{z}\!+\!4t_{J}^{}) \!+\! {\rm sign}(\Delta_{z}\!-\!4t_{J}^{}) \!-\! 2{\rm sign}(\Delta_{z}) \right]\!, \label{eq:C_0} 
\end{eqnarray} where ${\rm sign}(x)$ is the sign function.

\subsection{Numerical calculations}\label{3.2}

To numerically verify the analytical result in Eq.~\eqref{eq:C_0}, we derive the Berry curvature of the two bands of the Hamiltonian in Eq.~\eqref{eq:Hk} analytically 
\begin{eqnarray}
\Omega_{\pm,\bf k}^{xy} 
\!=\!\frac{\pm a^{2}t^{2}J_{u}}{2\left\{ \left[t\sin(k_{x}a)\right]^{2} \!+\! \left[t\sin(k_{y}a)\right]^{2} \!+\! J_{z}^{2} \!\right\}^{3/2}},\label{eq:Omega_xy}
\end{eqnarray}  where the signs $\pm$ correspond to the conduction and valence bands, respectively, 
and $J_{u}\!=\!2t_{J}^{}\left[\cos(k_{x}a) \!-\! \cos(k_{y}a)\right] \!-\! \Delta_{z}\cos(k_{x}a)\cos(k_{y}a)$.
Furthermore, the Chern number of the valence band is then obtained by numerically integrating the Berry curvature over the first Brillouin zone:
\begin{eqnarray}
C\!=\!-\int_{\rm B.Z.}\frac{d^{2}{\bf k}}{2\pi}\Omega_{-,\bf k}^{xy},\label{eq:C}
\end{eqnarray} where the Fermi energy $E_{F}$ is assumed to lie inside the bulk gap. 

As shown in Fig.~\ref{fig:phase}(b), we numerically evaluate the Chern number $C$ [Eq.~\eqref{eq:C}] of the valence band in the gapped regime at $E_{F}\!=\!0$ as a function of $t_{J}^{}$ and $\Delta_z$, obtaining the corresponding topological phase diagram. The system exhibits three distinct phases: a topological phase with $C\!=\!+1$ (red), a topological phase with $C\!=\!-1$ (blue), and a topologically trivial phase with $C\!=\!0$ (gray). The numerical results are in excellent agreement with the analytical result in Eq.~\eqref{eq:C_0}. In the following, we focus on the topologically nontrivial phase with $C\!=\!+1$.

\section{Joint density of states}\label{4}

To investigate the Van Hove singularity in the frequency domain, we calculate the joint density of states (JDOS). The JDOS characterizes the number of available electron--hole pairs between the valence and conduction bands that can participate in an interband optical transition with a photon of a given energy. It therefore provides information about the optical absorption spectrum and the characteristic frequencies associated with the electronic band structure. The JDOS is defined as the density of available interband transitions per unit energy interval and can be calculated as~\cite{mojarro2021optical,xiong2023optical}
\begin{eqnarray}
{\rm JDOS}(\omega)\!=\!\frac{1}{S}\sum_{\bf k}\delta\!\left[\hbar\omega\!-\!(\varepsilon_{+{\bf k}}^{}\!-\!\varepsilon_{-{\bf k}}^{})\right]\!,
\end{eqnarray}
where $S$ denotes the area of the sample, $\hbar\!=\!h/(2\pi)$ is the reduced Planck constant with the Planck constant $h$, and $\varepsilon_{\pm{\bf k}}^{}$ denote the eigenvalues for the model Hamiltonian.

\subsection{$t_{J}^{}\!\neq\!0$}\label{4.1}

In the presence of a finite $d$-wave altermagnetic term, $t_{J}^{}\!\neq\!0$, the characteristic interband transition frequencies at the high-symmetry points and Van Hove singularity are 
\begin{equation}\label{eq:omega_tJ}
\left\{
\begin{aligned}
&\omega_{\Gamma}^{}\!=\!\omega_{\textrm{M}}^{}\!=\!\omega_{0}^{}\!=\!\frac{2}{\hbar}|\Delta_{z}|,\\
&\omega_{\textrm{X}}^{}\!=\!\frac{2}{\hbar}|\Delta_{z}\!+\!4t_{J}^{}|,\\
&\omega_{\textrm{Y}}^{}\!=\!\frac{2}{\hbar}|\Delta_{z}\!-\!4t_{J}^{}|,\\
&\omega_{\textrm{V}}^{}\!=\!\frac{2}{\hbar}|\varepsilon_{\textrm{V}}^{}|.
\end{aligned}
\right.
\end{equation} Here, $\omega_{0}^{}$, $\omega_{\textrm{X}}^{}$, and $\omega_{\textrm{Y}}^{}$ denote the interband transition frequencies at the high-symmetry points $\Gamma$ (or $\textrm{M}$), $\textrm{X}$, and $\textrm{Y}$, respectively, while $\omega_{\textrm{V}}^{}$ is the frequency associated with the Van Hove singularity at energy $|\varepsilon_{\textrm{V}}^{}|$.

The Van Hove singularities correspond to stationary points of the energy dispersion in Eq.~\eqref{eq:Ek}, which satisfy $\partial\varepsilon_{\pm\bf k}^{}/\partial {\bf k}\!=\! 0$~\cite{xiong2023optical}.
Solving this equation under the condition $|t|\!<\!\sqrt{8t_{J}^{2} \!-\! \Delta_{z}^{2}/2}$ with $|\Delta_{z}|\!<\!|4t_{J}^{}|$, yields
\begin{eqnarray}
|\varepsilon_{\textrm{V}}^{}|&\!=\!&|t|\sqrt{2 \!+\! \frac{\Delta_z^2}{t^2 \!-\! 8t_J^2}},
\end{eqnarray} 
with the corresponding momentum-space location
$(\arccos[ 2t_{J}^{}\Delta_{z}/(t^{2} \!-\! 8t_{J}^{2})]/a,
\arccos[ -2t_{J}^{}\Delta_{z}/(t^{2} \!-\! 8t_{J}^{2})]/a)$.
Thus, the Van Hove singularity originates from stationary points whose energy and momentum locations depend explicitly on the model parameters. In particular, the corresponding characteristic frequency $\omega_{\textrm{V}}^{}$ can be tuned by varying the altermagnetic order $t_J^{}$, spin-orbit coupling strength $t$, and Zeeman splitting $\Delta_z$, providing a tunable spectroscopic signature of the underlying band structure.

\subsection{$t_{J}^{}\!=\!0$}\label{4.2}

In the absence of the $d$-wave altermagnetic term, i.e., $t_{J}^{}\!=\!0$, the characteristic frequencies reduce to
\begin{equation}\label{eq:omega_tJ_0}
\left\{
\begin{aligned}
&\omega_{\Gamma}^{}\!=\!\omega_{\textrm{M}}^{}\!=\!\omega_{\textrm{Y}}^{}\!=\!\omega_{\textrm{X}}^{}\!=\!\omega_{0}^{}\!=\!\frac{2}{\hbar}|\Delta_{z}|,\\
&\omega_{\textrm{V}}^{}\!=\!\frac{2}{\hbar}|\varepsilon_{\textrm{V}}^{}|,\\
&\omega_{c}^{}\!=\!\frac{2}{\hbar}|\varepsilon_{c}^{}|.
\end{aligned}
\right.
\end{equation}
Here, $\omega_{0}^{}$ denotes the characteristic interband transition frequency at the high-symmetry points $\Gamma$, $\textrm{M}$, $\textrm{X}$, and $\textrm{Y}$, while $\omega_{\textrm{V}}^{}$ is associated with the Van Hove singularity at energy $|\varepsilon_{\textrm{V}}^{}|$. The cutoff frequency $\omega_{c}^{}$ corresponds to the maximum interband transition energy $2|\varepsilon_{c}^{}|$, above which the JDOS vanishes.

In this case, solving the stationary-point condition $\partial\varepsilon_{\pm\bf k}^{}/\partial{\bf k}\!=\! 0$ yields the Van Hove singularity at
$|\varepsilon_{\textrm{V}}^{}|\!=\!\sqrt{t^2\!+\!\Delta_z^2}$, with the corresponding momentum-space location
$(0,\pm\pi/(2a))$ or $(\pm\pi/(2a), 0)$.
The maximum interband transition energy is attained at
$|\varepsilon_{c}^{}|\!=\!\sqrt{2t^2\!+\!\Delta_z^2}$,
corresponding to
$(\pm\pi/(2a),\pm\pi/(2a))$ or $(\pm\pi/(2a),\mp\pi/(2a))$.
Thus, in the absence of $d$-wave altermagnetism, the four high-symmetry points share the same gap frequency $\omega_{0}^{}$, while the Van Hove singularity and the upper transition cutoff are determined solely by the spin-orbit coupling strength $t$ and the Zeeman splitting $\Delta_z$.

\subsection{Numerical Results}\label{4.3}

\begin{figure}[htpb]
\centering
\includegraphics[width=\columnwidth]{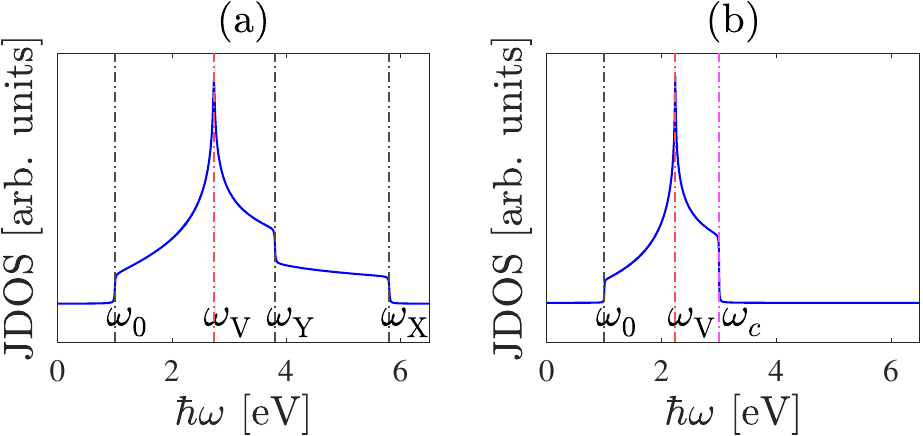}
\caption{Joint density of states (JDOS) as a function of the photon energy $\hbar\omega$ for (a) $t_{J}^{}\!=\!0.6$ eV and (b) $t_{J}^{}\!=\!0$. The other parameters are $t\!=\!1$ eV, $\Delta_{z}\!=\!-0.5$ eV, and $a\!=\!1$ nm.} \label{fig:JDOS_TB}
\end{figure}

In Figs.~\ref{fig:JDOS_TB}(a) and \ref{fig:JDOS_TB}(b), we plot the JDOS as a function of the photon energy $\hbar\omega$ for finite and vanishing $d$-wave altermagnetic coupling, respectively. The characteristic frequencies discussed above are indicated by the vertical dot-dashed lines.

When the photon frequency reaches the threshold frequency $\omega_0^{}$, indicated by the leftmost black dot-dashed line in Fig.~\ref{fig:JDOS_TB}, the photon energy $\hbar\omega_0^{}$ matches the minimum direct band gap, allowing interband transitions from the valence band to the conduction band. Consequently, the JDOS exhibits an abrupt onset at $\omega_0^{}$, as shown in Figs.~\ref{fig:JDOS_TB}(a) and \ref{fig:JDOS_TB}(b). As the frequency increases, the number of available interband transition channels generally increases, leading to an overall increase in the JDOS. A pronounced peak emerges at $\omega_{\textrm{V}}^{}$, indicated by the red dot-dashed line, which is associated with the Van Hove singularity. This feature can be understood in terms of the group velocity,
${\bf v}_{\pm}\!=\!\partial\varepsilon_{\pm{\bf k}}^{}/(\hbar\partial {\bf k})$.
At the Van Hove singularity, the group velocity vanishes, i.e., ${\bf v}_{\pm}\!=\!0$, indicating a stationary point of the band dispersion. As a result, electronic states become strongly concentrated within a narrow energy range, leading to an enhanced JDOS and giving rise to the pronounced peak at $\omega_{\textrm{V}}^{}$. Therefore, the peak at $\omega_{\textrm{V}}^{}$ provides a characteristic spectroscopic signature associated with the Van Hove singularity.

Beyond $\omega_{\textrm{V}}^{}$, the JDOS gradually decreases as the number of available interband transition channels is reduced. For finite $t_J^{}$, the $d$-wave altermagnetic term lifts the equivalence between the high-symmetry points $\textrm{X}$ and $\textrm{Y}$, introducing additional characteristic transition energies $\hbar\omega_{\textrm{X}}^{}$ and $\hbar\omega_{\textrm{Y}}^{}$. Accordingly, distinct changes in the JDOS occur at these frequencies, as shown in Fig.~\ref{fig:JDOS_TB}(a). These features reflect the redistribution of available interband transition channels induced by the anisotropic altermagnetic band structure.

For $t_J^{}\!=\!0$, the JDOS eventually vanishes above the cutoff frequency $\omega_c^{}$, as shown in Fig.~\ref{fig:JDOS_TB}(b). This frequency corresponds to the maximum interband transition energy, $2|\varepsilon_c^{}|$. Owing to the bounded lattice dispersion, when $\hbar\omega\!>\!2|\varepsilon_c^{}|$, no momentum states satisfy the energy-conservation condition for an interband transition. Consequently, the JDOS drops to zero above $\omega_c^{}$.

Overall, the characteristic features of the JDOS, including its threshold onset at $\omega_{0}^{}$, pronounced peak at the Van Hove frequency $\omega_{\textrm{V}}^{}$, and high-frequency cutoff at $\omega_{c}^{}$, provide clear signatures of the underlying band structure. In particular, the additional features at $\omega_{\textrm{X}}^{}$ and $\omega_{\textrm{Y}}^{}$ for finite $t_J^{}$ demonstrate how $d$-wave altermagnetism modifies the optical transition spectrum through its anisotropic band splitting.

\section{Optical Conductivities}\label{5}

In this section, we present the numerical results for the longitudinal and transverse optical conductivities and discuss their characteristic features.
At zero temperature, the optical conductivity $\sigma_{\alpha\beta}(\omega)$ can be evaluated using the Kubo formula~\cite{hu2022signature}
\begin{eqnarray}
\sigma_{\alpha\beta}(\omega)
&\!=\!&-\mathrm{i}\frac{e^{2}}{h}\sum_{m,n}\int \frac{d^{2}{\bf k}}{2\pi}\frac{\langle m|\partial_{\alpha}\hat{\cal H}|n\rangle\langle n|\partial_{\beta}\hat{\cal H}|m\rangle}{\varepsilon_{m\bf k}^{} \!-\! \varepsilon_{n\bf k}^{} \!+\! \hbar\omega  \!+\! \mathrm{i}\eta} \nonumber\\
&&\times\frac{\Theta(E_{F}\!-\!\varepsilon_{m\bf k}^{}) \!-\! \Theta(E_{F}\!-\!\varepsilon_{n\bf k}^{})}{\varepsilon_{m\bf k}^{} \!-\! \varepsilon_{n\bf k}^{}}, \label{eq:OH_T0_ab}
\end{eqnarray}
where $\alpha,\beta\!=\!x,y$ denote Cartesian coordinates, $\partial_{\alpha}\!\equiv\!\partial/(\partial k_{\alpha})$, $|n\rangle$ denotes an eigenstate of the Hamiltonian, $-e$ is the electron charge, $\eta$ is a positive infinitesimal, and $\Theta(x)$ is the Heaviside step function. Here, the summation runs over pairs of states $m$ and $n$ connected by optically allowed interband transitions.

\subsection{Longitudinal optical conductivity}\label{5.1}

\begin{figure}[htpb]
\centering
\includegraphics[width=\columnwidth]{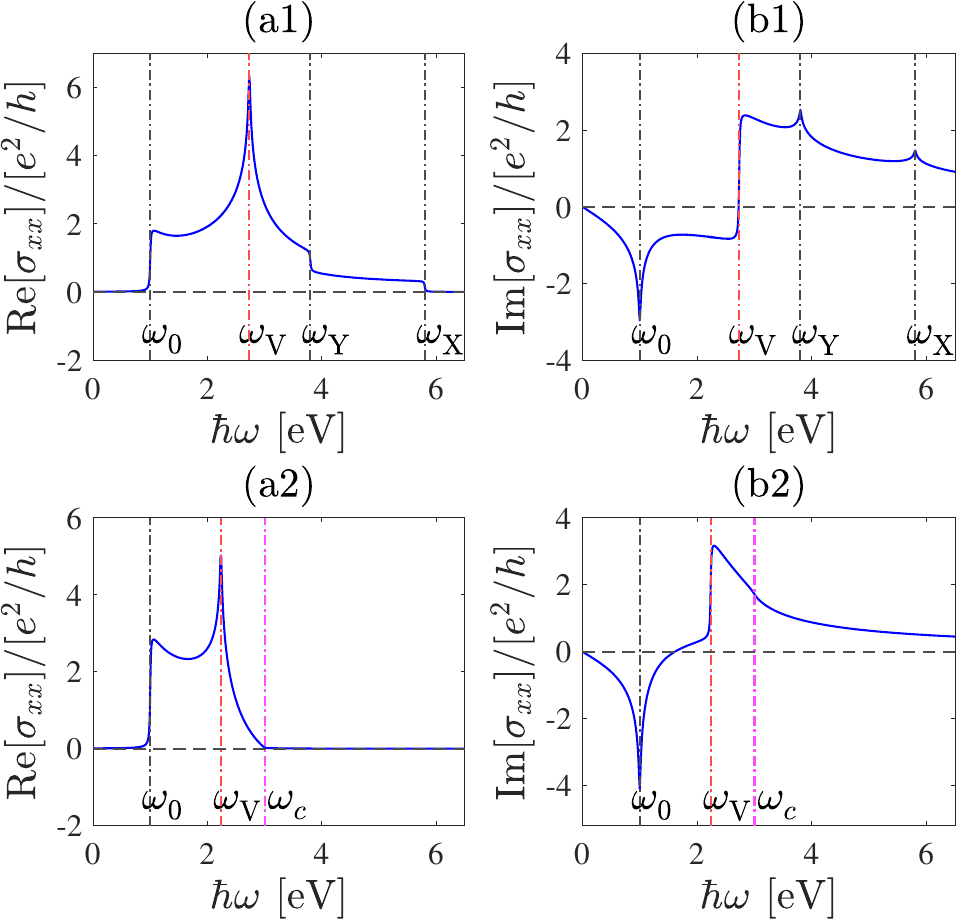}
\caption{Longitudinal optical conductivity $\sigma_{xx}$ [Eq.~\eqref{eq:OH_T0_ab}] as a function of the photon energy $\hbar\omega$ in the presence and absence of the $d$-wave altermagnetic coupling.
Upper row: (a1) real part and (b1) imaginary part of $\sigma_{xx}$ for $t_{J}^{}\!=\!0.6$ eV.
Lower row: (a2) real part and (b2) imaginary part of $\sigma_{xx}$ for $t_{J}^{}\!=\!0$.
The other parameters are $E_{F}\!=\!0$, $t\!=\!1$ eV, $\Delta_{z}\!=\!-0.5$ eV, and $a\!=\!1$ nm.} \label{fig:OH_TB_xx}
\end{figure}

In Figs.~\ref{fig:OH_TB_xx}(a1) and \ref{fig:OH_TB_xx}(a2), we plot the real part of the longitudinal optical conductivity, ${\rm Re}[\sigma_{xx}(\omega)]$, as a function of the photon energy $\hbar\omega$ in the presence and absence of the $d$-wave altermagnetic coupling, respectively. In the zero-frequency limit, ${\rm Re}[\sigma_{xx}(0)]$ vanishes in both cases, as shown in Figs.~\ref{fig:OH_TB_xx}(a1) and \ref{fig:OH_TB_xx}(a2).

For frequencies below the threshold, $\omega\!<\!\omega_0^{}$, the real part of the longitudinal optical conductivity vanishes because no interband optical transitions are energetically allowed. In this regime, the interband transitions are Pauli blocked, and the photon energy is insufficient to bridge the minimum direct band gap. Once the critical frequency $\omega\!=\!\omega_0^{}$ is reached, interband transitions become allowed, leading to the abrupt onset of ${\rm Re}[\sigma_{xx}(\omega)]$ and the emergence of a step-like feature. As the frequency increases further, ${\rm Re}[\sigma_{xx}(\omega)]$ develops a pronounced peak around $\omega_{\textrm{V}}^{}$, which originates from the enhanced optical transition phase space associated with the Van Hove singularity. This behavior is consistent with the pronounced peak in the JDOS at $\omega_{\textrm{V}}^{}$ for both finite and vanishing $t_J^{}$, as shown in both Figs.~\ref{fig:OH_TB_xx}(a1) and \ref{fig:OH_TB_xx}(a2). 

For finite $t_J^{}$, ${\rm Re}[\sigma_{xx}(\omega)]$ decreases for $\omega\!>\!\omega_{\textrm{V}}^{}$ and exhibits two pronounced drops at $\omega_{\textrm{Y}}^{}$ and $\omega_{\textrm{X}}^{}$, respectively, as shown in Fig.~\ref{fig:OH_TB_xx}(a1). These features originate from the corresponding changes in the JDOS and reflect the redistribution of available interband transition channels induced by the $d$-wave altermagnetic splitting.

For $t_J^{}\!=\!0$, in the frequency range $\omega_{\textrm{V}}^{}\!<\!\omega\!<\!\omega_{c}^{}$, ${\rm Re}[\sigma_{xx}(\omega)]$ gradually decreases, mainly reflecting the reduction in the available interband transition phase space and, consequently, the decrease in the JDOS. For $\omega\!>\!\omega_c^{}$, ${\rm Re}[\sigma_{xx}(\omega)]$ vanishes because the JDOS is zero, as shown in Fig.~\ref{fig:OH_TB_xx}(a2).

According to the Kramers--Kronig relations~\cite{hu2022signature,xiong2023optical}, the real and imaginary parts of the optical conductivity are related by causality. We therefore plot the imaginary part, ${\rm Im}[\sigma_{xx}(\omega)]$, as a function of the optical frequency in Figs.~\ref{fig:OH_TB_xx}(b1) and \ref{fig:OH_TB_xx}(b2). The corresponding structures in the imaginary part provide a complementary characterization of the optical response and are generally consistent with the characteristic features observed in the real part. Interestingly, however, the feature associated with the cutoff frequency $\omega_c^{}$ in ${\rm Re}[\sigma_{xx}(\omega)]$ is absent in ${\rm Im}[\sigma_{xx}(\omega)]$, as shown in Fig.~\ref{fig:OH_TB_xx}(b2). This behavior indicates that the imaginary part is insensitive to the abrupt vanishing of the JDOS at $\omega_c^{}$ and is instead determined by the dispersive contributions from the optical transitions over the entire frequency range.

In the above discussion, we focus on the longitudinal conductivity $\sigma_{xx}(\omega)$. We have also verified that similar behavior is obtained for $\sigma_{yy}(\omega)$. The corresponding numerical results for $\sigma_{yy}(\omega)$ are presented in Fig.~S1 of Section~SIII of the Supplemental Material~\cite{SuppMat}.

\subsection{Transverse optical conductivity}\label{5.2}

\begin{figure}[htpb]
\centering
\includegraphics[width=\columnwidth]{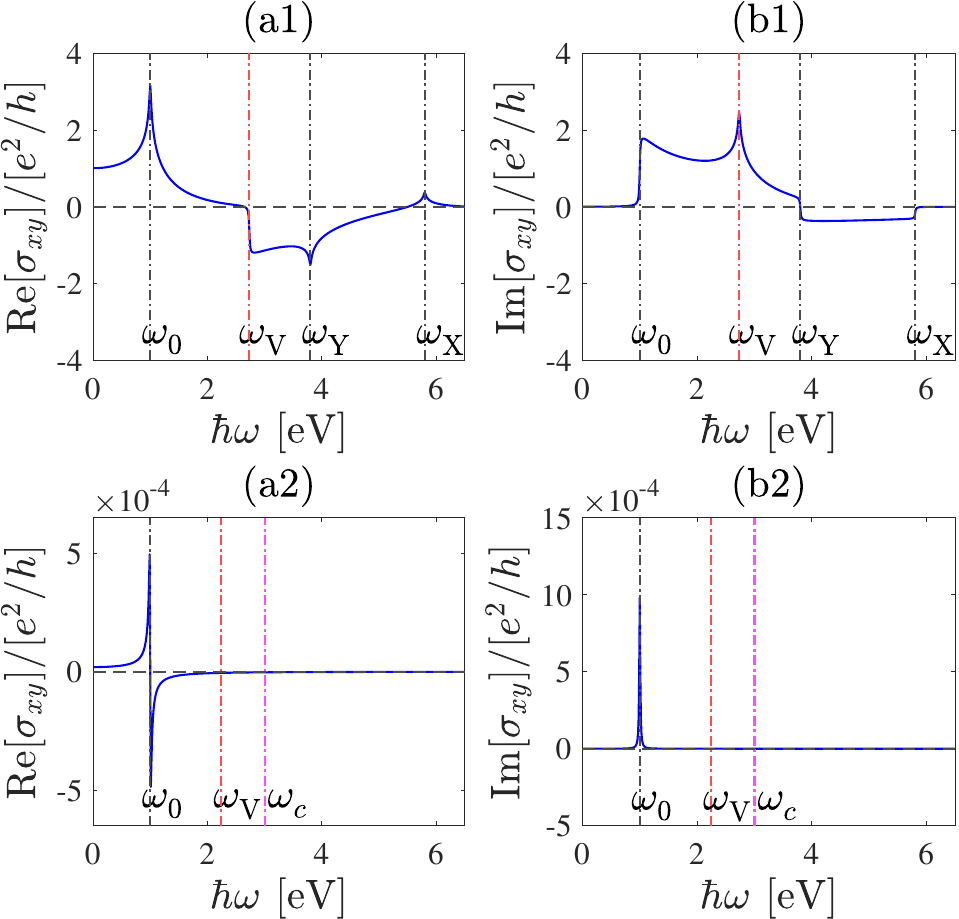}
\caption{Transverse optical conductivity $\sigma_{xy}$ [Eq.~\eqref{eq:OH_T0_ab}] as a function of the photon energy $\hbar\omega$ in the presence and absence of the $d$-wave altermagnetic coupling.
Upper row: (a1) real part and (b1) imaginary part of $\sigma_{xy}$ for $t_{J}^{}\!=\!0.6$ eV.
Lower row: (a2) real part and (b2) imaginary part of $\sigma_{xy}$ for $t_{J}^{}\!=\!0$.
The other parameters are the same as those used in Fig.~\ref{fig:OH_TB_xx}.} \label{fig:OH_TB_xy}
\end{figure}

In Fig.~\ref{fig:OH_TB_xy}(a1) and Fig.~\ref{fig:OH_TB_xy}(a2), we plot the real part of the transverse optical conductivity, ${\rm Re}[\sigma_{xy}(\omega)]$, as a function of the photon energy $\hbar\omega$ in the presence and absence of the $d$-wave altermagnetic coupling, respectively. For finite $t_J^{}$, the zero-frequency transverse conductivity ${\rm Re}[\sigma_{xy}(0)]/[e^2/h]$ is quantized in the topologically nontrivial regime, as shown in Fig.~\ref{fig:OH_TB_xy}(a1). This quantization is consistent with the nonzero Chern number of the occupied band and provides a direct manifestation of the underlying topological phase. In contrast, for $t_{J}^{}\!=\!0$, the system is topologically trivial, and ${\rm Re}[\sigma_{xy}(0)]$ nearly vanishes, as shown in Fig.~\ref{fig:OH_TB_xy}(a2). 

For frequencies below the threshold, $\omega\!<\!\omega_{0}^{}$, ${\rm Re}[\sigma_{xy}(\omega)]$ increases with increasing frequency in both cases. For finite $t_{J}^{}$, once the critical frequency $\omega\!=\!\omega_{0}^{}$ is reached, interband transitions become allowed, resulting in a pronounced peak in ${\rm Re}[\sigma_{xy}(\omega)]$. As the frequency increases further, ${\rm Re}[\sigma_{xy}(\omega)]$ decreases and exhibits a pronounced dip at $\omega_{\textrm{V}}^{}$, which is associated with the Van Hove singularity. The corresponding feature is closely related to the enhanced optical transition phase space at the Van Hove singularity. For $\omega\!>\!\omega_{\textrm{V}}^{}$, ${\rm Re}[\sigma_{xy}(\omega)]$ exhibits a sharp valley at $\omega_{\textrm{Y}}^{}$ followed by a pronounced peak at $\omega_{\textrm{X}}^{}$. These characteristic structures originate from the corresponding changes in the JDOS and reflect the anisotropic interband transition spectrum induced by the $d$-wave altermagnetic coupling, as shown in Fig.~\ref{fig:OH_TB_xy}(a1).

For $t_{J}^{}\!=\!0$, once the critical frequency $\omega\!=\!\omega_{0}^{}$ is reached, ${\rm Re}[\sigma_{xy}(\omega)]$ exhibits a pronounced singular behavior, as shown in Fig.~\ref{fig:OH_TB_xy}(a2). For $\omega\!>\!\omega_{0}^{}$, the magnitude $\left|{\rm Re}[\sigma_{xy}(\omega)]\right|$ gradually decreases, primarily reflecting the reduction in the available interband transition phase space. In contrast to the finite-$t_{J}^{}$ case, however, no distinct feature associated with the Van Hove singularity is observed in ${\rm Re}[\sigma_{xy}(\omega)]$. This absence indicates that the prominent Van Hove signature in the transverse optical conductivity is induced by the $d$-wave altermagnetic coupling.

Since the imaginary and real parts of the optical conductivity are related through the Kramers--Kronig relations, the corresponding characteristic features are also reflected in ${\rm Im}[\sigma_{xy}(\omega)]$, as shown in Fig.~\ref{fig:OH_TB_xy}(b1) and Fig.~\ref{fig:OH_TB_xy}(b2).

\section{Magneto-optical effects}\label{6}

Experimentally, the optical conductivity can be probed through magneto-optical measurements~\cite{tse2010giant,tse2011magneto,lei2023kerr,kargarian2015theory}. In particular, the Faraday and Kerr rotations provide sensitive probes of the optical and topological properties of thin films of altermagnetic topological systems.

When an optical beam is incident on the quantum material, the wave vectors of the incident, reflected, and transmitted beams can be written as~\cite{xiong2023optical}
\begin{eqnarray}
{\bf k}_{l}^{}&\!=\!&(k_{l}^{}\sin\theta_{l}\cos\phi, k_{l}^{}\sin\theta_{l}\sin\phi, \eta_{l}^{}k_{l}^{}\cos\theta_{l}), 
\end{eqnarray} where $l\!=\!i,r,t$, $\eta_{i}\!=\!\eta_{t}\!=\!-1$, $\eta_{r}\!=\!1$, $\phi$ is the azimuthal angle of the plane of incidence, while $\theta_i$, $\theta_r$, and $\theta_t$ denote the incident, reflected, and transmitted angles, respectively. 

The electric field vector of the optical beam can be decomposed into the $s$- and $p$-polarized components, which are perpendicular and parallel to the plane of incidence, respectively:
\begin{eqnarray}
{\bf E}_{l}&\!=\!&\left(E_{l}^{s}{\bf e}_{l}^{s} \!+\! E_{l}^{p}{\bf e}_{l}^{p}\right)e^{\mathrm{i}({\bf k}_{l}^{}\cdot{\bf r}-\omega_{l}^{}t)}, \label{eq:Ej0}
\end{eqnarray} where $l\!=\!i,r,t$, and the corresponding unit polarization vectors are defined as
${\bf e}_{l}^{s}\!=\!(\sin\phi)\hat{\mathbf{x}} \!-\! (\cos\phi)\hat{\mathbf{y}}$ and ${\bf e}_{l}^{p}\!=\!{\bf e}_{l}^{s}\!\times\!\hat{\bf k}_{l}^{}$ with $\hat{\bf k}_{l}^{}\!=\!{\bf k}_{l}^{}/k_{l}^{}$.

Here, we consider only the free-standing case, i.e., $n_{i}\!=\!n_{r}\!=\!n_{t}\!=\!1$, and focus on normally incident light with $\theta_{i}\!=\!0$ and $\phi\!=\!\pi/2$. By solving Maxwell's equations with the appropriate boundary conditions, the Faraday and Kerr rotation angles, $\theta_{\rm F}^{s/p}$ and $\theta_{\rm K}^{s/p}$, can be expressed as~\cite{xiong2023optical}
\begin{eqnarray}
\theta_{\rm F/K}^{s/p}\!=\!\frac{1}{2}\arctan\!\left\{ \frac{2{\rm Re}\!\left[ \chi_{\rm F/K}^{s/p} \right]}{1 \!-\! \big|\chi_{\rm F/K}^{s/p}\big|^{2}} \right\}\!,\label{eq:angles}
\end{eqnarray} where the auxiliary quantities characterizing the polarization rotation of the transmitted and reflected light are given by
$\chi_{\rm F}^{s}\!=\!\tilde{\alpha}\tilde{\sigma}_{xy}/(1 \!+\! \tilde{\alpha}\tilde{\sigma}_{yy})$, 
$\chi_{\rm F}^{p}\!=\!\tilde{\alpha}\tilde{\sigma}_{xy}/(1 \!+\! \tilde{\alpha}\tilde{\sigma}_{xx})$,
$\chi_{\rm K}^{s}\!=\!\tilde{\sigma}_{xy}/[\tilde{\alpha}(\tilde{\sigma}_{xy}^{2} \!+\! \tilde{\sigma}_{xx}\tilde{\sigma}_{yy}) \!+\! \tilde{\sigma}_{xx}]$, and 
$\chi_{\rm K}^{p}\!=\!\tilde{\sigma}_{xy}/[\tilde{\alpha}(\tilde{\sigma}_{xy}^{2} \!+\! \tilde{\sigma}_{xx}\tilde{\sigma}_{yy}) \!+\! \tilde{\sigma}_{yy}]$.
Here, $\tilde{\sigma}_{\alpha\beta}\!=\!\sigma_{\alpha\beta}/(e^{2}/h)$ ($\alpha,\beta\!=\!x,y$) denotes the dimensionless optical conductivity, and $\tilde{\alpha}\!=\![e^2/(2h)]\sqrt{\mu_{0}^{}/\varepsilon_{0}^{}}\!=\!1/137$ is the vacuum fine-structure constant. The quantities $\varepsilon_{0}^{}$ and $\mu_{0}^{}$ denote the vacuum permittivity and permeability, respectively. A detailed derivation of the Faraday and Kerr angles in Eq.~\eqref{eq:angles} is provided in Section SIV of the Supplemental Material~\cite{SuppMat}.

\begin{figure}[htpb]
\centering
\includegraphics[width=\columnwidth]{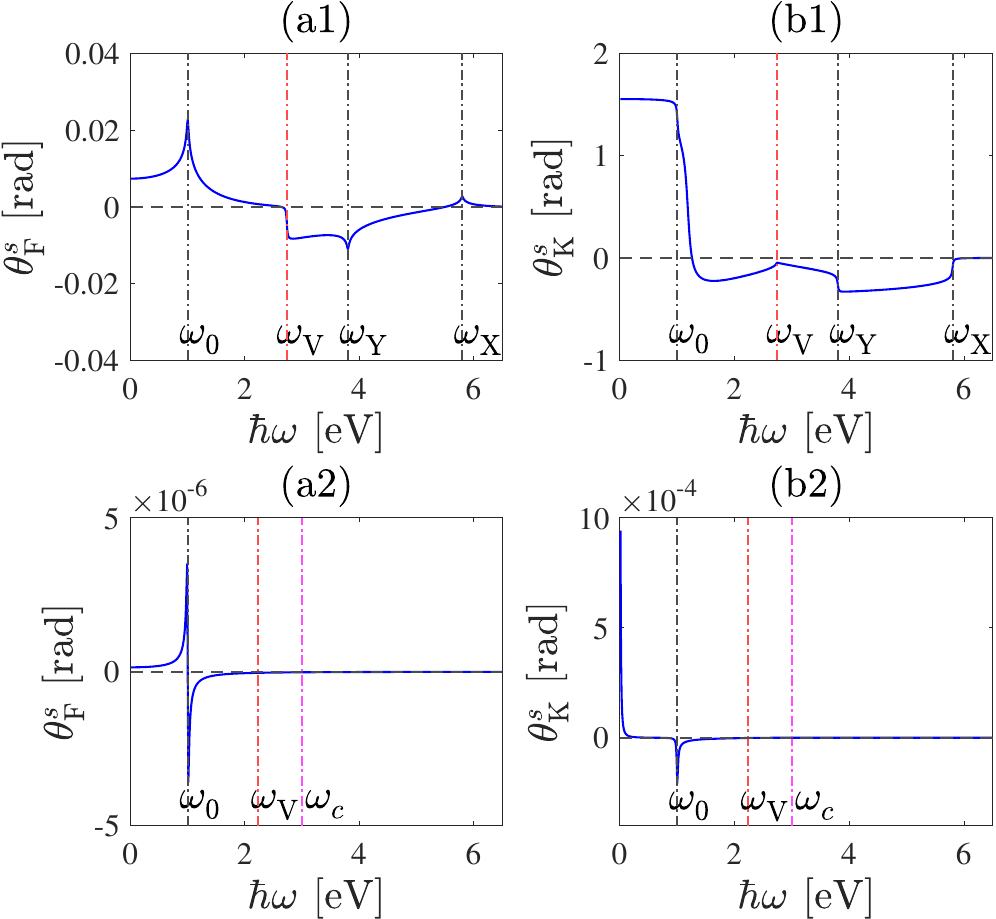}
\caption{The $s$-polarized Faraday rotation angle $\theta_{\rm F}^{s}$ and Kerr rotation angle $\theta_{\rm K}^{s}$ as functions of the photon energy $\hbar\omega$, with and without the $d$-wave altermagnet. Upper row: (a1) Faraday rotation angle $\theta_{\rm F}^{s}$ and (b1) Kerr rotation angle $\theta_{\rm K}^{s}$ for $t_{J}^{}\!=\!0.6$ eV. Lower row: (a2) Faraday rotation angle $\theta_{\rm F}^{s}$ and (b2) Kerr rotation angle $\theta_{\rm K}^{s}$ for $t_{J}^{}\!=\!0$. Here, we consider only the free-standing case, i.e., $n_{i}\!=\!n_{r}\!=\!n_{t}\!=\!1$, and set $\theta_{i}\!=\!0$ and $\phi\!=\!\pi/2$. The other parameters are the same as those used in Fig.~\ref{fig:OH_TB_xx}.} \label{fig:Faraday_s_Kerr_s}
\end{figure}

In Fig.~\ref{fig:Faraday_s_Kerr_s}, we plot the $s$-polarized component of Faraday and Kerr rotation angles as functions of the incident light energy $\hbar\omega$ for the normally incident light with $\theta_{i}\!=\!0$ and $\phi\!=\!\pi/2$. 
Interestingly, both the Faraday and Kerr rotation angles exhibit characteristic spectral features that can be traced back to the underlying optical conductivity. In particular, the frequency dependence of the Faraday rotation angle in Fig.~\ref{fig:Faraday_s_Kerr_s}(a1) closely resembles that of the real part of the transverse optical conductivity, ${\rm Re}[\sigma_{xy}(\omega)]$, shown in Fig.~\ref{fig:OH_TB_xy}(a1). In contrast, the $s$-polarized Kerr rotation angle in Fig.~\ref{fig:Faraday_s_Kerr_s}(b1) approximately follows the frequency dependence of the imaginary part, ${\rm Im}[\sigma_{xy}(\omega)]$, shown in Fig.~\ref{fig:OH_TB_xy}(b1). These results demonstrate that the characteristic frequencies appearing in the optical conductivity are directly reflected in the Faraday and Kerr spectra, providing experimentally accessible signatures for extracting information about the underlying electronic and altermagnetic parameters of topological materials through magneto-optical spectroscopy.

In the above discussion, we focus on the $s$-polarized component. We have also verified that similar characteristic features are present in the $p$-polarized Faraday and Kerr rotation angles. The corresponding numerical results for $\theta_{\rm F}^{p}$ and $\theta_{\rm K}^{p}$ are presented in Fig.~S2 of Section~SIV of the Supplemental Material~\cite{SuppMat}.

\section{Summary}\label{7}

We study the optical and magneto-optical signatures of Dirac points and Van Hove singularities in a two-dimensional $d$-wave altermagnetic topological system with spin-orbit coupling and Zeeman splitting. The system hosts gapped Dirac points at $\Gamma$, $\textrm{M}$, $\textrm{X}$, and $\textrm{Y}$, whose gap frequencies produce distinct signatures in the JDOS and optical conductivities. In particular, the JDOS exhibits characteristic kinks at the Dirac gaps and pronounced peaks at Van Hove singularities, whose positions are tunable by the altermagnetic order. These features are transferred to the longitudinal and transverse optical conductivities, with complementary structures in their real and imaginary parts. Specifically, the Van Hove signatures in the transverse conductivity are absent without $d$-wave altermagnetism, demonstrating an altermagnet-induced optical fingerprint of the Van Hove singularity. The corresponding characteristic features are further manifested in the frequency-dependent Faraday and Kerr rotations. Our results highlight optical and magneto-optical spectroscopy as effective probes of Dirac gaps, Van Hove singularities, and altermagnetic order in topological systems.


\begin{acknowledgments}
R.C. acknowledges support from the National Natural Science Foundation of China (Grants No. 12304195, No. 12674198, and No. U25D8012), the Chutian Scholars Program in Hubei Province, the Key Project of Hubei Provincial Department of Education (under Grant No. D20241004), the Hubei Provincial Natural Science Foundation (Grant No. 2025AFA081), the Wuhan City Key R\&D Program (under Grant No. 2025050602030069), and the Original Seed Program of Hubei University.
This work is supported by the Guangdong Provincial Quantum Science Strategic Initiative (Grant No. GDZX2401001).
F.Q. acknowledges support from the Jiangsu Specially-Appointed Professor Program in Jiangsu Province and the Doctoral Research Start-Up Fund of Jiangsu University of Science and Technology. Xiao-Bin Qiang  acknowledges the program of China Scholarship Council (Grant No. 202508440272).
\end{acknowledgments}

\section*{Data Availability}
The data are available from the authors upon reasonable request.

%
%
%
\twocolumngrid
\bibliography{references_Altermagnets_Optical_Hall}

\end{document}